\documentclass[twocolumn,preprintnumbers,amsmath,amssymb]{revtex4-2}
\usepackage{graphicx}
\usepackage{dcolumn}
\usepackage{bm}

\begin{document}

\title{
Laser focusing equivalence behavior of mass-limited targets
in laser particle acceleration
}

\author{Toshimasa~Morita}

\affiliation{Kansai Institute for Photon Science,
National Institutes for Quantum Science and Technology,
8-1-7 Umemidai, Kizugawa, Kyoto 619-0215, Japan}


\begin{abstract}
The generation of $200$ MeV class protons by irradiating a $25$ J laser pulse
onto a water target using three-dimensional particle-in-cell simulation
is shown.
Two types of targets---foil and disk---are evaluated and compared.
Disks, which are mass-limited targets, produce ions with higher energy than
those produced by foils.
This enhancement is attributed to laser focusing equivalent effects in
mass-limited targets during the acceleration process.
In addition, the disk diameter that generates the maximum energy protons is
theoretically derived, which shows good agreement with the simulation result.
\end{abstract}

\maketitle

\section{Introduction}

Laser ion acceleration, in which lasers are employed to accelerate ions, offers
the potential for compact and cost-effective ion accelerators \cite{BWE,DNP}.
Recent advancements in compact laser system performance have rendered laser ion
accelerators feasible.
However, the ion energy currently obtained by laser acceleration remains lower
than that needed for practical applications such as
hadron therapy \cite{WDB,HGK}, which requires above $200$ MeV protons.
Therefore, further increasing the energy of the generated ions has become
a key research topic \cite{BWP,DL,PRK,SPJ,Toncian,KKQ,MEBKY,TM1,TM2}.
Two approaches exist for obtaining high-energy ions via laser acceleration: the
first involves enhancing laser performance (intensity and energy), and the
second focuses on improving target design.
Nevertheless, enhancing laser performance typically requires larger and more
expensive laser systems, thereby undermining the objective of developing
compact and cost-effective ion accelerators.
Therefore, this study adopted the latter approach, which revealed fluctuations
in ion energy produced by different targets and the reasons for this difference,
to generate ions with higher energy by using the optimal target.
Thus, this paper presents the conditions for efficiently generating
high-energy ($\sim 200$ MeV) protons.

The target used for proton generation can be improved in two ways: selecting
an appropriate material and optimizing the shape.
Hydrogen-rich targets are effective in producing high-energy protons,
as suggested in Ref. \onlinecite{TM2}, which utilized CH$_2$.
As CH$_2$ and H$_2$O exhibit nearly identical electron densities when
fully ionized and have comparable charge-to-mass ratios for C and O,
H$_2$O is anticipated to generate high-energy protons as efficiently as CH$_2$.
Moreover, H$_2$O offers practical advantages: it is readily available,
can be shaped into various forms due to its liquid state at room temperature,
and can be easily solidified into ice.
Consequently, water (H$_2$O) was selected as the target material for this study.

Two target shapes were examined: a foil and a disk.
Foils are straightforward to fabricate due to their simple shape,
while disks have the potential to produce ions with higher energy \cite{TM1}.
The thickness of the foil and both the thickness and diameter of the disk
were varied in the simulations to assess the resulting proton energies and
the underlying causes of any differences.

The remainder of this paper is organized as follows.
Section \ref{para} presents the simulation parameters.
The results for the foil and disk targets are presented and discussed
in Section \ref{resul}.
The analytical considerations are detailed in Section \ref{theory},
and the key findings of this study are summarized in Section \ref{con}.

\section{Simulation model} \label{para}

Simulations were performed using a three-dimensional (3D) parallelized
electromagnetic code, based on the particle-in-cell (PIC) method \cite{CBL}.
The following parameters were applied in the simulations:

An idealized model, in which a Gaussian linear polarized laser pulse was
normally incident on a target represented by a collisionless plasma, was used.
The laser pulse with the dimensionless amplitude $a_0=q_eE/m_{e}\omega c=72$,
corresponding to the laser peak intensity, $I$, of $1\times 10^{22}$ W/cm$^{2}$,
had a $30$ fs full width at half maximum (FWHM) duration and
was focused on a spot of size $2.5 \mu$m (FWHM).
The laser had a power, $P$, of $0.8$ PW and energy, $\mathcal{E}_\mathrm{las}$,
of $25$ J; the laser wavelength was $\lambda = 0.8 \mu$m.
The foil and disk cases differed solely in the terms of the shape of the target,
with all other conditions remaining the same.

The laser propagation direction was set along the $X$ direction,
and the electric field was oriented in the $Y$ direction.
The boundary conditions for the particles and fields were periodic in the
transverse $(Y,Z)$ directions,
and absorbing at the boundaries of the computation box along the $X$ axis.
The number of grid cells was $5300 \times 3456 \times 3456$ along the $X$, $Y$,
and $Z$ axes, respectively.
Correspondingly,
the simulation box size was $95\mu$m$\times 62\mu$m$\times 62\mu$m.

The target was positioned with its surface lying in the $YZ-$plane.
Foil targets were defined in the entire simulation box in the $Y$ and
$Z$ directions.
The center of the disk target was at $Y/2$, $Z/2$.
The laser-irradiated surfaces of the foil and disk were placed at $X=32 \mu$m,
and the center of the laser pulse was located $16 \mu$m in front of it.

Foil thicknesses in the range of 0.05--0.2 $\mu$m, and disk thicknesses and disk
diameters in the ranges of 0.1--0.3 $\mu$m and 1.0--10 $\mu$m, respectively,
were considered.
These ranges facilitate the generation of high-energy protons under the
above-mentioned laser conditions \cite{TM2}.
Simulations were performed at several points within each parameter range.
For a foil thickness of $0.1 \mu$m, the total number of quasiparticles was
$2\times 10^{10}$.
The number of quasiparticles varied proportionally to the thickness.
For example, if the target thickness was doubled ($0.2\mu$m),
the number of particles also doubled.
The total number of quasiparticles in a disk with a thickness of $0.2\mu$m and
diameter of $3.0\mu$m was $8\times 10^7$; as in the case of the foil,
the number of particles obtained with disks of different thicknesses was
a multiple of the disk thickness; similarly, the number of particles obtained
with disks of different diameters was a multiple of the disk area.
The ionization state of an oxygen ion was assumed to be $Z_{i}=+8$.
The electron density was $n_{e}=3\times 10^{23}$ cm$^{-3}$,
i.e., $n_{e}=190 n_\mathrm{c}$, where $n_c$ denotes the critical density.
The proton density was $n_{e}/5$, and the oxygen ion density was $n_{e}/10$.

An $xyz-$coordinates system was used throughout the text and figures.
The origin of the coordinate system was located at the center of the
laser-irradiated surface of the initial target,
and the directions of the $x,y,$ and $z$ axes were the same as those of the
$X,Y$ and $Z$ axes, respectively.
Therefore, the $x$ axis denoted the direction perpendicular to the target
surface, and the $y$ and $z$ axes were parallel to the target surface.

\section{Simulation results} \label{resul}

\begin{figure}[tbp]
\includegraphics[clip,width=\hsize,bb=43 35 547 351]{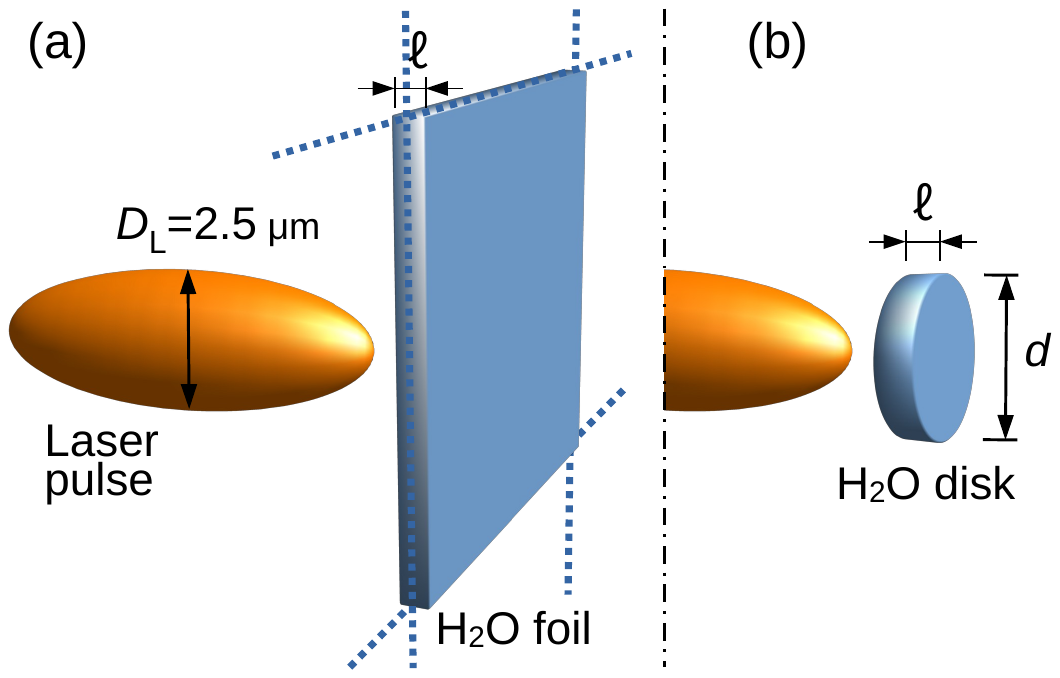}
\caption{
Laser pulse (spot diameter: $2.5\mu$m; energy: $25$ J) is irradiated onto
H$_2$O foil (a) and disk (b) targets.
The thickness $\ell$ of the foil and the thickness $\ell$ and diameter $d$ of
the disk are varied to investigate the characteristics of the generated ions.
}
\label{fig:fig-1}
\end{figure}

The simulation results for foil and disk targets composed of H$_2$O are
presented in this section.
These results highlight variations in the energy and spatial distribution of
ions and electrons due to differences in target shape (foil versus disk),
thickness $\ell$, and disk diameter $d$.

It was demonstrated that radiation pressure acceleration (RPA) becomes
the dominant mechanism for $I/n_e \ell > 500$ W/electron \cite{TM3}.
Specifically, $I/n_e \ell = 1100$ W/electron (which exceeds $500$ W/electron),
even in the thickest target case ($0.3\mu$m),
where its value was the smallest in our simulations.
Therefore,
the acceleration scheme in the early stage of ion acceleration is RPA.

Targets made of hydrogen-rich materials, such as H$_2$O and CH$_2$,
experience a strong Coulomb explosion, which is the main acceleration scheme
in the later stage of the acceleration process.
Thus, the acceleration scheme in this study comprise the RPA and
the subsequent Coulomb explosion \cite{TM3}.

If the target is excessively thick, the efficiency of RPA diminishes,
leading to a reduction in the maximum ion energy \cite{TM3}.
Conversely, if the target is too thin, since the number of electrons
per unit surface area of the target decreases with the thickness,
even if all electrons within the target are accelerated,
the number of electrons interacting with ions decreases.
Thus, the effect of ion acceleration diminishes,
and the generated ion energy decreases.
Therefore, under certain laser conditions, an optimal target thickness exists.

Additionally, when the disk target diameter is significantly smaller than
the laser spot diameter, $D_L$, the portion of the laser that
passes through the area beyond the circumference of the disk (the area where
the disk is not present) increases, i.e., the energy of the laser pulse
absorbed by the disk target decreases; consequently,
the maximum generated ion energy decreases.
Conversely, as its diameter increases, the behavior of the disk becomes
closer to that of a foil, leading to a decrease in the maximum ion energy.
Thus, an optimal disk diameter exists for certain laser conditions.
Accordingly, we investigated the maximum proton energy generated at multiple
thicknesses, $\ell$, for foils and disks, and diameter, $d$,
for disks (Fig. \ref{fig:fig-1}).

\begin{figure}[tbp]
\includegraphics[clip,width=\hsize,bb=27 36 560 400]{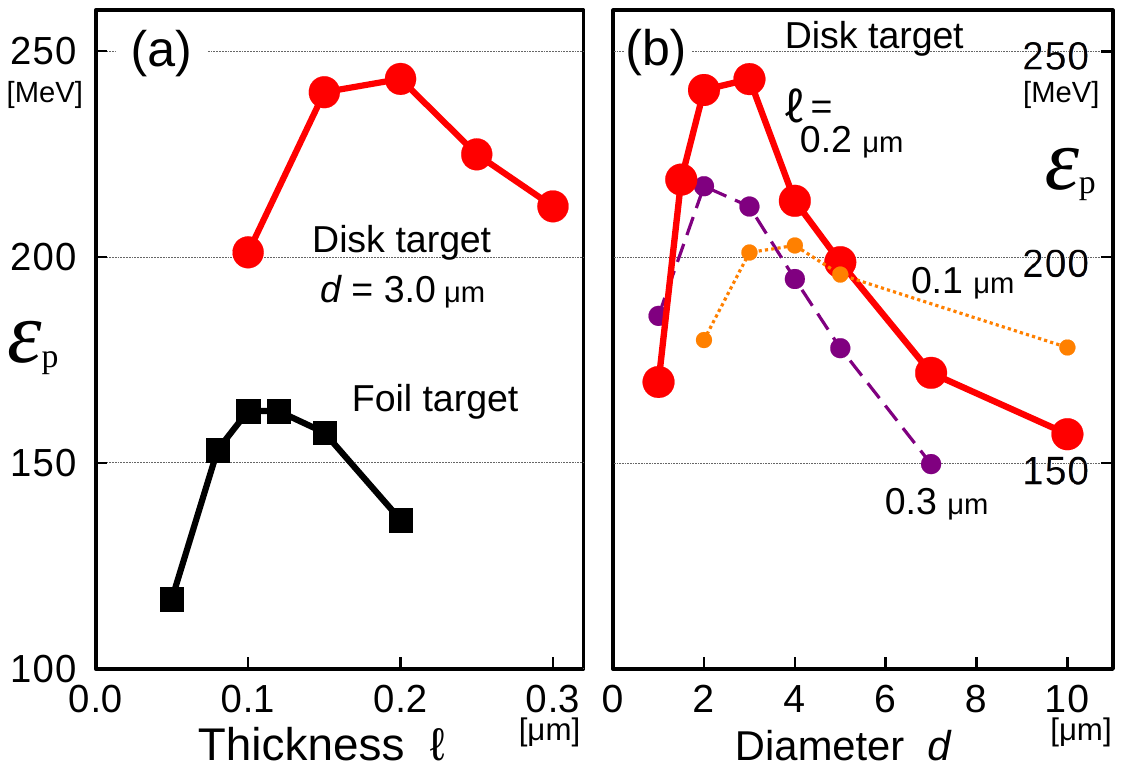}
\caption{
(a) Maximum proton energy, $\mathcal{E}_p$, for each target thickness, $\ell$,
of the foil and disk.
The disk produces protons with approximately 1.5 times higher energy than
that obtained with the foil.
An optimal target thickness exists for generating high-energy protons,
which is greater for the disk than for the foil.
This suggests the occurrence of a phenomenon similar to laser focusing in disks.
(b) Maximum proton energy, $\mathcal{E}_p$, at each disk diameter, $d$.
The results for disk thicknesses of 0.1, 0.2, and 0.3 $\mu$m are shown.
There exists an optimum disk diameter (i.e., $3\mu$m at thickness
$\ell = 0.2\mu$m) at which high-energy protons are produced.
The optimal disk diameter is $1.2 \times D_L$, where $D_L$ is
the laser spot diameter (FWHM).
}
\label{fig:fig-2}
\end{figure}

Figure \ref{fig:fig-2} shows the maximum proton energies, $\mathcal{E}_p$,
achieved for various target thicknesses and disk diameters.
Each curve exhibits a peak, indicating the existence of an optimal target
thickness as well as diameter in the case of disks for generating high-energy
protons.
Figure \ref{fig:fig-2}(a) shows $\mathcal{E}_p$ for different target thicknesses
of the foils (black square) and disks (with a diameter $d$ of $3\mu$m; red
circle); $d=3\mu$m was selected because it produces the highest
$\mathcal{E}_p$ (see Fig. \ref{fig:fig-2}(b)).
As shown in Fig. \ref{fig:fig-2}(a), the optimal thicknesses are $0.1\mu$m for
the foil and $0.2\mu$m for the disk, producing protons with energies of
160 and 240 MeV, respectively.
The disk target generates protons with 1.5 times higher energy than
that obtained with the foil target, and the optimal disk thickness is twice
the optimal foil thickness.
In general, in laser ion acceleration, a larger number of electrons are expelled
per unit area of the target surface, resulting in higher ion energies.
To expel more electrons from a unit area of the target surface, an increased
energy per unit area is required,
necessitating the use of high-intensity lasers.
In addition,
the higher the laser intensity, the higher the target thickness that produces
high-energy ions.
The fact that the optimal thickness of the disk is greater than that of the foil
suggests that a phenomenon similar to the increased laser intensity (achieved
by laser focusing) is occurring in the disks.
This phenomenon is explored in detail below.

Figure \ref{fig:fig-2}(b) shows the maximum proton energy, $\mathcal{E}_p$,
as a function of disk diameter for thicknesses of $0.1\mu$m (orange),
$0.2\mu$m (red), and $0.3\mu$m (purple).
For each thickness, the maximum $\mathcal{E}_p$ is attained for disk diameters
in the range of 2 to 4 $\mu$m.
Outside of this diameter range, $\mathcal{E}_p$ declines sharply.
The optimal disk diameter for producing high-energy protons is the largest
for a disk thickness of $0.1\mu$m, followed by those of $0.2\mu$m and
$0.3\mu$m thicknesses.
The maximum $\mathcal{E}_p$ for each thickness and diameter is observed at
$d=3\mu$m and $\ell=0.2\mu$m.
The maximum $\mathcal{E}_p$ is produced when the disk diameter is 1.2 times
the laser spot size (FWHM), i.e., diameter $D_L$.
The larger the laser spot diameter, the larger the disk diameter that produces
the high-energy ions. 

The variation in the target thickness as well as the variation in the disk
diameter significantly influence the generation of high-energy protons.
Therefore, to produce high-energy protons, targets must have appropriate
thicknesses as well as appropriate disk diameters for the laser conditions.

\begin{figure}[tbp]
\includegraphics[clip,width=0.9\hsize,bb=28 38 563 435]{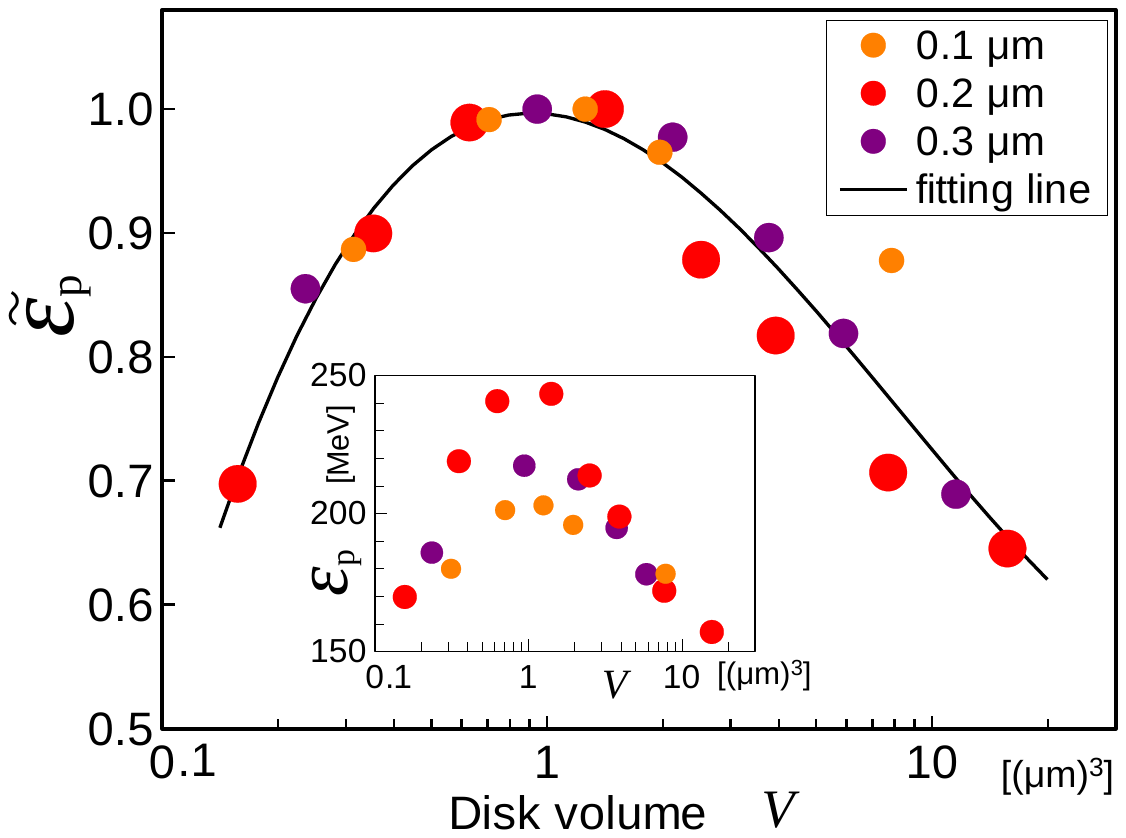}
\caption{
Relationship between the disk target volume and obtained maximum proton energy,
$\tilde{\mathcal{E}}_p$ (normalized by the maximum value at each thickness).
$\tilde{\mathcal{E}}_p$ is the maximum at a volume of $\sim$1 $\mu$m$^3$,
regardless of the disk thickness.
Inset:
the vertical axis is the non-normalized proton energy, $\mathcal{E}_p$ [MeV].
}
\label{fig:fig-3}
\end{figure}

As shown in Fig. \ref{fig:fig-2}(b), for a thicker disk, the maximum
$\mathcal{E}_p$ value is obtained at a smaller diameter; that is,
the relationship between the thickness and diameter of a disk that produces
high-energy protons is such that the volume of the disk remains constant.
Figure \ref{fig:fig-3} shows the relationship between the disk volume and
$\mathcal{E}_p$ for the cases shown in Fig. \ref{fig:fig-2}(b).
Here, the values of $\mathcal{E}_p$ are normalized such that its
maximum value at each thickness is $1$; that is,
$\tilde{\mathcal{E}}_p=\mathcal{E}_p/\mathcal{E}_{p,max}$,
where $\mathcal{E}_{p,max}$ is the maximum value for each thickness.
The thicknesses of 0.1, 0.2, and 0.3 $\mu$m are depicted by orange, red,
and purple circles, respectively.
The solid black line is the least-squares-fit line for all points.
Regardless of disk thickness, the peak $\tilde{\mathcal{E}}_p$ occurs at
a volume of approximately 1 $\mu$m$^3$.
$\tilde{\mathcal{E}}_p$ decreases as the volume deviates from this optimal
value, with the curves exhibiting similar trends irrespective of thickness.
This is because the laser energy remains the same in all cases in this
study; therefore, the disk volume, i.e., the total number of electrons
that receive the energy as efficiently as possible, would be the same.

High-energy ions can be generated by selectively applying as much laser energy
as possible to as small a volume of the target as possible
(a small number of electrons).
The following approaches can be adopted to provide as much of the laser energy
as possible to the disk target: 1. Increase the disk diameter to minimize the
amount of laser surpassing the area occupied by the disk; 2. Increase the disk
thickness so that the amount of laser transmitted through the disk is minimized.
Therefore, the volume of the disk, $V$, will increase.
However, to selectively deliver laser energy to a minimal number of electrons,
$V$ must be minimized; i.e., both the disk diameter and its thickness should be
small.
These conflicting requirements necessitate an optimal volume
(diameter and thickness) for disk targets.
This volume is determined from the laser conditions
(intensity, pulse duration, and spot size), which were the same in all cases in
this study.
Therefore, the optimal disk volume $V$ for the generation of high-energy ions is
the same at 0.1, 0.2, and 0.3 $\mu$m.
Figure \ref{fig:fig-3} (inset) shows the non-normalized proton energy
$\mathcal{E}_p$ [MeV] along the vertical axis.
The maximum $\mathcal{E}_p$ varied with thickness; however, the peak occurred
at $V \approx$ 1 $\mu$m$^3$ for each thickness.

In the following sections, the detailed considerations for the $0.1\mu$m-thick
foil and the $0.2\mu$m-thick disk with a diameter of $3\mu$m,
which generate the maximum $\mathcal{E}_p$ for foil and disk targets,
respectively, are presented. 

\begin{figure*}[tbp]
\includegraphics[clip,width=0.9\hsize,bb=31 32 546 329]{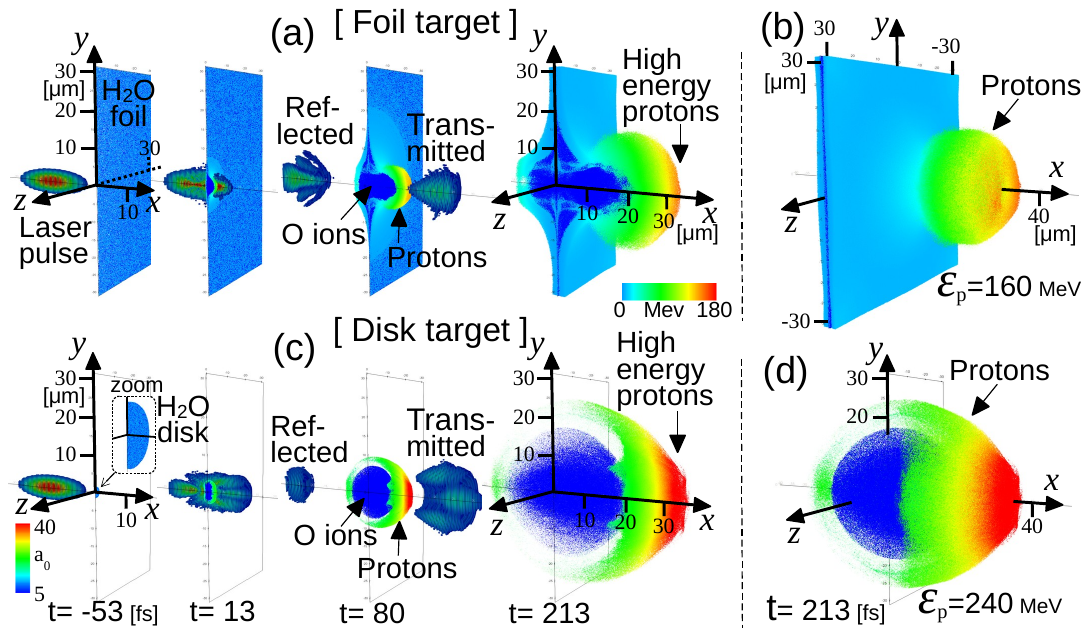}
\caption{
3D view of the particle distribution and electric field magnitude (isosurface
for $a_0=5$) at each time.
The laser pulse is normally incident on the H$_2$O foil and disk targets.
In (a,c), half of the electric field box and ion cloud have been excluded
to reveal the internal structure.
For protons, color corresponds to energy.
(a) Foil target with $\ell=0.1\mu$m and (b) overall view.
(c) Disk target with $d=3\mu$m, $l=0.2\mu$m, and (d) overall view.
}
\label{fig:fig-4}
\end{figure*}

Figure \ref{fig:fig-4} presents the particle distribution and electric field
magnitude for both the $0.1\mu$m-thick foil (a,b) and $0.2\mu$m-thick disk
with a diameter of $3\mu$m (c,d).
Figure \ref{fig:fig-4}(a) and \ref{fig:fig-4}(c) show the foil and disk cases,
respectively, at identical time points.
Here, half of the electric field box and ion distribution, that is, the $z>0$
region, has been excluded to reveal the internal structure.
The protons are classified by color in terms of their energy,
with red indicating high energy and light blue indicating low energy,
while the oxygen, O, ions are depicted in blue.
The instant when the center of the laser pulse, where the laser intensity $I$
is the strongest, reaches the laser-irradiated surface of the initial target
is assumed as $t=0$.
Therefore,
more than half of the laser pulse does not interact with the target when $t<0$,
and more than half of the interaction is complete at $t \ge 0$.
The simulation start time is $t=-53$ fs.
The initial shapes of the laser pulse and target are shown at $t=-53$ fs.
The laser pulse is defined on the $-x$ side of the target and propagates
in the $+x$ direction.
At $t=13$ fs, the laser pulse undergoes strong interactions with the target,
and at this point, approximately half of the laser pulse has interacted with
the target.
At this time, in both the foil and disk, the O ions and protons separate and
form two layers within the laser irradiation region.
At $t=80$ fs, the interaction between the laser pulse and target ends, and the
laser pulse is partly reflected from and partly transmitted through the target.
The O ions are distributed near the center of the ion cloud, and the protons
are distributed around them.
The simulation end time is $t=213$ fs; the overall states at this time are shown
in Fig. \ref{fig:fig-4}(b,d).
In the foil, the O ions are distributed near the center of the ion cloud that
is generated around the laser-irradiated area, and numerous protons are
distributed spherically on its +x side, with a maximum energy of 160 MeV at
the tip of the +x side.
In the disk, the ion cloud is spherical and is distributed separately into
protons and O ions, with the O ions near the center and the protons distributed
on its $+x$ side.
The proton with a maximum energy of 240 MeV is generated at the tip.
The disk target produces high-energy protons with 1.5 times higher energy than
that of the protons obtained with the foil target.
Thus,
disk targets produce high-energy protons more efficiently than foil targets.

\begin{figure}[tbp]
\includegraphics[clip,width=0.9\hsize,bb=37 34 563 395]{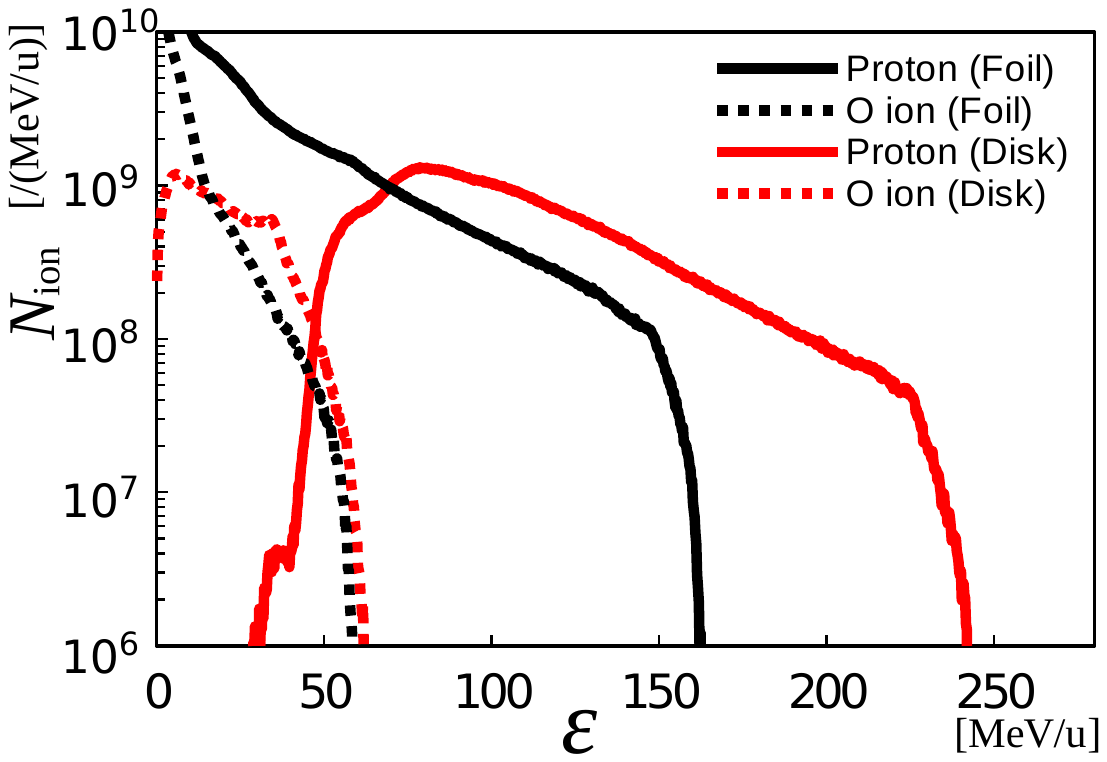}
\caption{
Energy spectrum of ions traveling in the $+x$ direction,
obtained in the simulation at $t=213$ fs.
The disk produces higher-energy protons than the foil, and in the high-energy
region ($\gtrsim 70$ MeV), the number of generated protons is greater than
that obtained with the foil.
}
\label{fig:fig-5}
\end{figure}

The energy spectra of the protons and O ions in the foil and disk cases at
$t=213$ fs are shown in Fig. \ref{fig:fig-5}.
Here, only the ions that travel to the $+x$ side are considered, because
the ions are received and used by some device installed behind the target,
i.e., in the $+x$ direction.
The results of the foil and disk cases are shown by black and red lines,
respectively,
and those of protons by a solid line; the O ions are indicated by a dotted line.
In the disk case, 200 MeV-class protons are generated with significantly higher
energy than that in the foil case, and in sufficient numbers
($\sim 10^8$ protons/MeV) for practical applications \cite{ESI,BEE}.
Although the energy levels differ between the foil and disk, the curves in the
relatively high-energy range, that is, $\mathcal{E} \gtrsim 100$ MeV for the
disk and $\mathcal{E} \gtrsim 40$ MeV for the foil, have similar shapes.
This is because the distributions of the high-energy protons are similar for the
foil and disk, that is, both are spherical, as shown in Fig. \ref{fig:fig-4}.
The protons in the low-energy range ($\mathcal{E} < 30$ MeV) are rarely present
in the disk case but are abundant in the foil case.
These are the protons in the area surrounding the laser-irradiated part of
the foil.
However, for O ions in the relatively high-energy range
($\mathcal{E}>15$ MeV/u), both the energy and number of ions are nearly
identical between the foil and disk cases.

\begin{figure}[tbp]
\includegraphics[clip,width=0.9\hsize,bb=36 30 561 402]{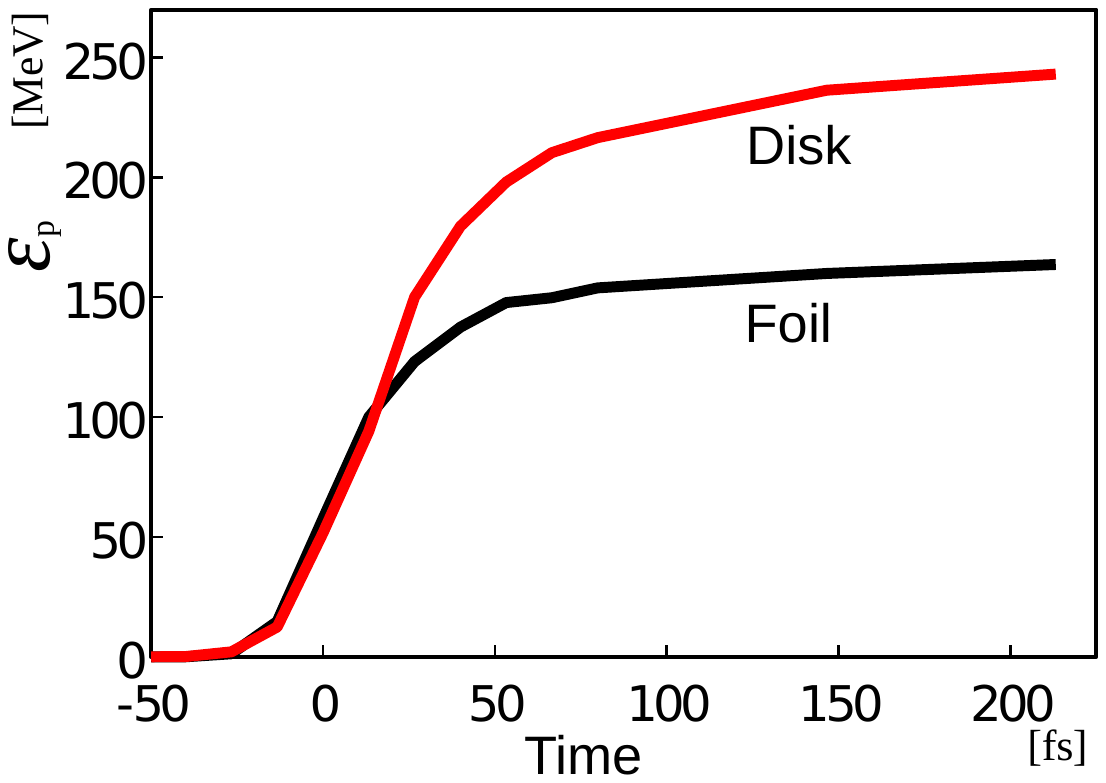}
\caption{
Maximum proton energy, $\mathcal{E}_p$, as a function of time.
For $t \lesssim 15$ fs, the energies of protons produced by the disk and foil
targets are almost the same, but after $t \sim$15 fs, the energy of the
protons produced by the disk target gradually becomes higher than that of the
protons produced by the foil target,
eventually becoming approximately 1.5 times higher.
}
\label{fig:fig-6}
\end{figure}

Figure \ref{fig:fig-6} shows $\mathcal{E}_p$ for the foil and disk as
a function of time.
For times $t \lesssim 15$ fs, $\mathcal{E}_p$ is almost the same for the foil
and disk.
However, for $t \gtrsim 15$ fs, $\mathcal{E}_p$ in the disk case gradually
becomes higher than that in the foil case.
The laser pulse width is 30 fs (FWHM), and the center of the laser pulse is
on the target surface at $t=0$; thus, $t=15$ fs is the backward time when
the laser intensity is halved, indicating that most of the laser pulse
irradiation on the target has concluded by this point.
This shows that, from this time onward, the disk accelerates the protons more
strongly than the foil.
The reason is that an effect equivalent to laser focusing comes into play in the
disks, as shown below.
In the following text,
we explain why disks produce higher energy ions than foils.

\begin{figure*}[tbp]
\includegraphics[clip,width=\hsize,bb=33 33 632 170]{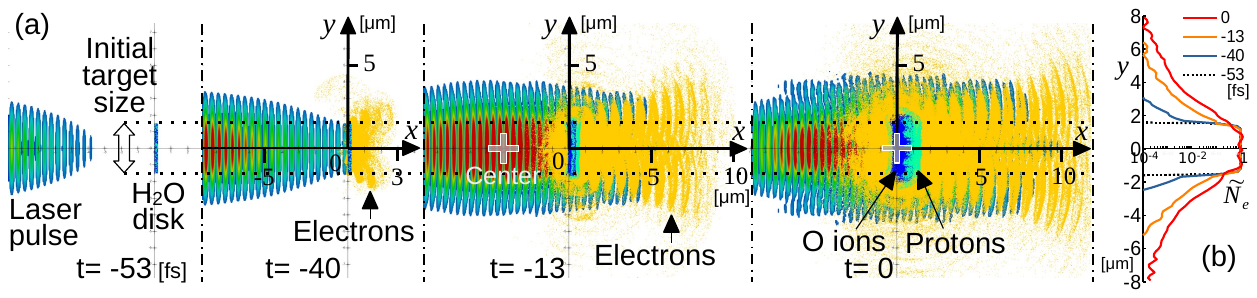}
\caption{
(a) 2D projection of the disk target case of the section at $z=0$ shown as
viewed along the $z$ axis.
Distribution of particles and laser pulse around the target in the early stage.
The electrons emitted by the target spread widely in the $y$ direction and
interact with the laser passing through the area beyond the circumference of
the disk, i.e., these electrons absorb the given laser energy.
This effect is equivalent to laser focusing.
Owing to RPA, numerous protons are accelerated toward the $+x$ direction and
form two layers with the oxygen ions.
(b) Number distribution of electrons in the $y$ direction at each time,
normalized by the value at $t=-53$ fs.
As time passes,
the electrons become more widely distributed in the $y$ direction. 
}
\label{fig:fig-7}
\end{figure*}

Figure \ref{fig:fig-7}(a) presents the distributions of ions and electrons
around the disk target and the laser pulse at early times ($t \le 0$).
The electrons are shown in yellow, protons in light blue,
and oxygen ions in blue.
Here, we show the cross section at $z=0$ in two dimensions (2D).
The state considered from the $z=0$ surface to a depth of $0.75\mu$m (half of
the disk radius) in the $z$ direction is viewed from a position at $z>0$.
This figure also shows the electrons that are not depicted in
Fig. \ref{fig:fig-4} and illustrates how the electrons pushed out from the
disk are distributed over a wide area.
The two black horizontal dotted lines in the figure indicate the positions of
the upper and lower ends of the initial disk target, and the laser pulse center
is marked by a white cross.
The electrons to be pushed out from the target at an early stage by the laser
pulse are distributed over a wider area than the ions, and this distribution
area expands over time.
In contrast, the protons and O ions have much smaller distribution areas,
and their variations with time are also smaller.
In Fig. \ref{fig:fig-7}(a), $t=-53$ fs is the initial state,
with the laser pulse has not yet reached the target.
At $t=-40$ fs, the center of the laser pulse is located $12\mu$m behind the
target; at this time, only the tip of the laser pulse has contacted the target
surface, but many electrons have already been expelled from the target.
However, the position of the ion distribution does not change significantly.
At $t=-13$ fs, the laser pulse center is positioned at $x=-4\mu$m,
not interacting with the target,
even though numerous electrons are distributed over a wide area.
At this time, the vertical ($y$ direction) distribution range of the electrons
extends to approximately three times the initial target size at around
$x \approx 7\mu$m.
These electrons absorb the energy of the laser pulse, which is about to reach,
as it passes through the area beyond the circumference of the disk (the area
above the upper black dotted line and below the lower black dotted line).
At this time, the distribution range of the ions (protons and O ions) expands
slightly in the $x$ direction but is almost the same as the initial range in
the $y$ direction.
Additionally, the protons appear on the $+x$ side, and the O ions are
distributed behind them (on the $-x$ side).
This is because many protons are strongly accelerated toward the $+x$ direction
by the RPA.
At $t=0$, the center of the laser pulse reaches the surface of the target on the
$-x$ side, i.e., half of the laser pulse completes interacting with the target
(it is partly transmitted through the target and partly reflected from it).
At this time, the distribution range of the electrons in the $y$ direction is
approximately two times the initial target size near $x=0$ (initial target
position) and approximately 3--4 times the initial target size on the $+x$
side ($x>5\mu$m).

Figure \ref{fig:fig-7}(b) depicts the distributions of the number of electrons
in the $y$ direction at each instant shown in Fig. \ref{fig:fig-7}(a).
Here, they are normalized so that the maximum value at the initial time
($t=-53$ fs) is 1.
The electrons become widely distributed in the $y$ direction as time passes.

A laser pulse with a spot diameter of $2.5\mu$m (FWHM) would appear to have
a large portion passing through the area beyond the disk with a diameter of
$3\mu$m.
However, in reality, the electrons in the disk are widely distributed in the
area beyond the circumference of the disk in the early stages and absorb the
energy of the laser passing through the area outside the disk,
as shown in Fig. \ref{fig:fig-7}.
These are the electrons that originally exist within the initial disk.
This implies that the part of the laser pulse passing through the area beyond
the disk is absorbed by the electrons that are originally inside the disk.
From a different perspective,
the laser pulse in the area beyond the disk heats the initial disk region,
which is similar to laser focusing.
This explains the generation of high-energy ions from mass-limited targets,
such as disks.

In the following text,
we demonstrate through additional considerations that the laser energy
outside the disk is absorbed by the electrons expelled from the disk.

\begin{figure}[tbp]
\includegraphics[clip,width=\hsize,bb=18 37 569 412]{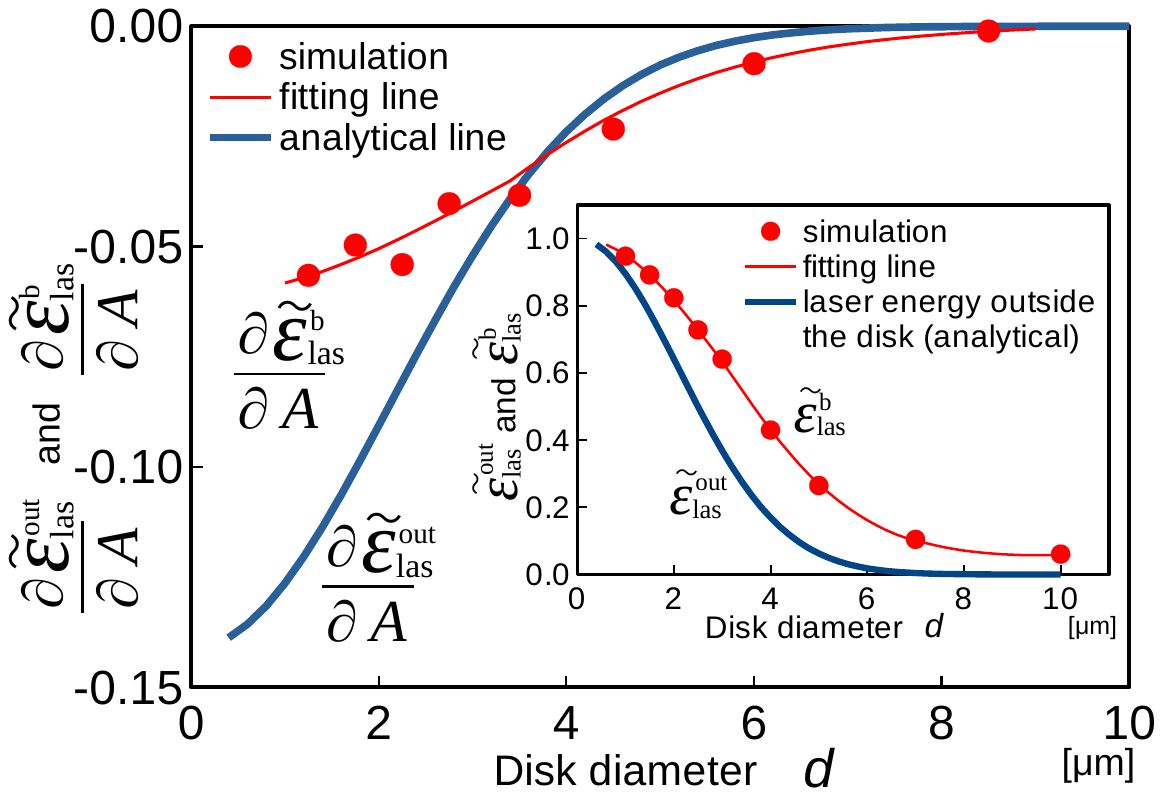}
\caption{
$\partial \tilde{\mathcal{E}}^\mathrm{b}_\mathrm{las}/\partial A$ and
$\partial \tilde{\mathcal{E}}^\mathrm{out}_\mathrm{las}/\partial A$
at each disk diameter $d$.
The red line is the fitting line.
$\tilde{\mathcal{E}}^\mathrm{b}_\mathrm{las}$ denotes the energy of the
electromagnetic wave arriving at the $+x$ boundary of the simulation box
in the simulation, and
$\tilde{\mathcal{E}}^\mathrm{out}_\mathrm{las}$ is the laser energy incident on
the area beyond the circumference of the initial disk target (area outside the
circle of diameter $d$) in the analysis.
Both are values normalized by the laser pulse energy, $\mathcal{E}_\mathrm{las}$.
$A$ is the disk surface area.
For a small disk diameter,
$\partial \tilde{\mathcal{E}}^\mathrm{b}_\mathrm{las}/\partial A >> \partial \tilde{\mathcal{E}}^\mathrm{out}_\mathrm{las}/\partial A$,
because the electrons pushed out from the disk increase its diameter.
Inset: $\tilde{\mathcal{E}}^\mathrm{b}_\mathrm{las}$ (simulation) and
$\tilde{\mathcal{E}}^\mathrm{out}_\mathrm{las}$ (analytical) at each disk
diameter.
Both lines show similar trends.
}
\label{fig:fig-8}
\end{figure}

As these expelled electrons absorb the laser energy passing outside the disk,
the energy of the laser arriving at the boundary on the $+x$ side of the
simulation box should be lower than expected.
We use this reasoning to depict the relationship between the laser energy
arriving at the boundary on the $+x$ side and disk diameter
(Fig. \ref{fig:fig-8}).
In Fig. \ref{fig:fig-8} (inset), the energy of the electromagnetic wave that
reaches the $+x$ boundary of the simulation box,
$\mathcal{E}^\mathrm{b}_\mathrm{las}$ (specifically, calculated as the sum of
the energies of the laser passing through the area beyond the circumference of
the disk, laser transmitted through the disk, and electromagnetic wave emitted
by the moving electrons) is shown by a red circle for each disk diameter
in the simulation.
The red line is their fitting line.
The analytical line (blue solid line) is the laser energy,
$\mathcal{E}^\mathrm{out}_\mathrm{las}$, in the area outside a circle of
diameter $d$ placed with its plane facing the laser (i.e., outside the
circumference of the initial disk target).
This is the energy of the laser pulse arriving at the boundary on the $+x$ side
under the assumption that the disk target retains its initial shape without
undergoing ionization or allowing laser transmission.
Here, both values are shown as their respective ratios to the energy of the
initial laser pulse, $\mathcal{E}_\mathrm{las}$, i.e.,
$\tilde{\mathcal{E}}^\mathrm{b}_\mathrm{las} = \mathcal{E}^\mathrm{b}_\mathrm{las}/\mathcal{E}_\mathrm{las}$ and $\tilde{\mathcal{E}}^\mathrm{out}_\mathrm{las}=\mathcal{E}^\mathrm{out}_\mathrm{las}/\mathcal{E}_\mathrm{las}$; $\mathcal{E}^\mathrm{out}_\mathrm{las}$
is given by
$\mathcal{E}^\mathrm{out}_\mathrm{las}=1-\tilde{\mathcal{E}}_\mathrm{R}=e^{-aR^2}$ from Eq. (\ref{r_ene})
in Section \ref{theory}.
The $\tilde{\mathcal{E}}^\mathrm{b}_\mathrm{las}$ and
$\tilde{\mathcal{E}}^\mathrm{out}_\mathrm{las}$ curves are similar.
However, for all diameters,
$\tilde{\mathcal{E}}^\mathrm{b}_\mathrm{las} > \tilde{\mathcal{E}}^\mathrm{out}_\mathrm{las}$,
because $\tilde{\mathcal{E}}^\mathrm{out}_\mathrm{las}$ does not account for
the energy of the laser transmitted through the target and energy of the
electromagnetic wave emitted by the moving electrons.
For a disk with smaller diameter, $\tilde{\mathcal{E}}^\mathrm{b}_\mathrm{las}$
is large, i.e., a large amount of the electromagnetic field energy reaches the
boundary on the $+x$ side.
As the diameter increases, $\tilde{\mathcal{E}}^\mathrm{b}_\mathrm{las}$
decreases significantly and eventually converges to a small constant value.
Because the diameter is sufficiently large at $d=10\mu$m,
$\tilde{\mathcal{E}}^\mathrm{b}_\mathrm{las}$ is the laser energy transmitted
through the target plus the energy of the electromagnetic field generated by
the motion of the electrons, which is very small at 0.06 ($6\%$ of the energy
of the laser pulse).
The differentials of $\tilde{\mathcal{E}}^\mathrm{b}_\mathrm{las}$ and
$\tilde{\mathcal{E}}^\mathrm{out}_\mathrm{las}$ with respect to the disk
surface area $A$,
$\partial \tilde{\mathcal{E}}^\mathrm{b}_\mathrm{las}/\partial A$ and
$\partial \tilde{\mathcal{E}}^\mathrm{out}_\mathrm{las}/\partial A$, at each
disk diameter are shown in Fig. \ref{fig:fig-8} by red circles and blue solid
line, respectively.
The red solid line is the fitting line.
The values are negative, that is,
$\partial \tilde{\mathcal{E}}^\mathrm{b}_\mathrm{las}/\partial A <0$ and
$\partial \tilde{\mathcal{E}}^\mathrm{out}_\mathrm{las}/\partial A <0$,
for all diameters.
This indicates that an increase in disk area ($\Delta A>0$), i.e., an increase
in disk diameter, reduces the laser energy reaching the boundary on the $+x$
side, whereas a decrease in disk area ($\Delta A<0$) increases that energy.
As the diameter becomes smaller, the values of
$\partial \tilde{\mathcal{E}}^\mathrm{b}_\mathrm{las}/\partial A$ and
$\partial \tilde{\mathcal{E}}^\mathrm{out}_\mathrm{las}/\partial A$
become smaller.
The simulation and analytical values are almost the same for $d \gtrsim 3\mu$m,
and both have a maximum value of $\sim 0$ for $d \gtrsim 6\mu$m.
In contrast, for $d \lesssim 3\mu$m, the simulation and analytical values
gradually show a greater difference as the diameter decreases, and for small
diameters, the simulation value is considerably larger than the analytical
value.
This indicates that for $d \lesssim 3\mu$m, the amount of laser passing through
the area beyond the disk does not decrease significantly in the simulation
even when the disk diameter is reduced.
This is because electrons expelled from the disk counter the effect of the
smaller diameter, that is, the effect of the smaller disk diameter is reduced
because the laser energy in the area beyond the disk, through which the laser
normally passes, is absorbed by the electrons that have been expelled into
that area.

\begin{figure}[tbp]
\includegraphics[clip,width=\hsize,bb=28 38 560 400]{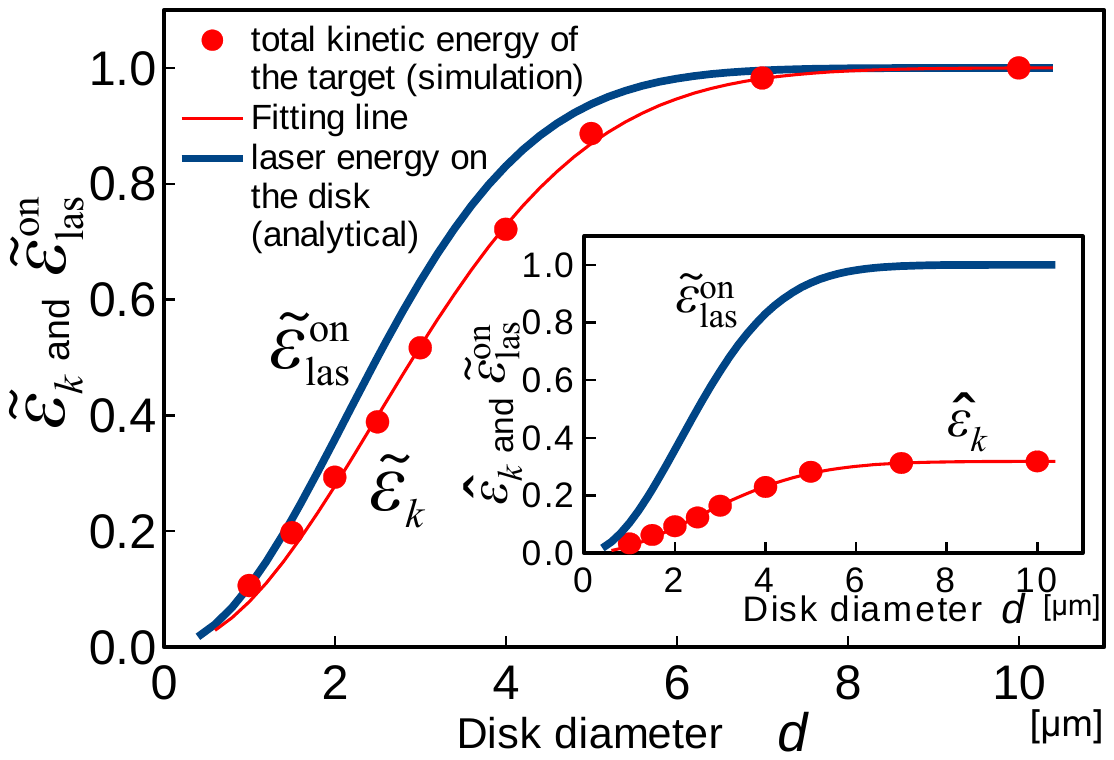}
\caption{
Total kinetic energy, $\tilde{\mathcal{E}}_k$, gained by the disk target at each
diameter, and laser energy incident on the initial disk surface (circular area
of diameter $d$),
$\tilde{\mathcal{E}}^\mathrm{on}_\mathrm{las}$, both normalized by their
respective maximum values.
Both lines depict an upward trend until a certain diameter ($d \gtrsim 7\mu$m)
and then become stable.
Inset: $\tilde{\mathcal{E}}^\mathrm{on}_\mathrm{las}$ can also be regarded as
normalized by the total laser energy, $\mathcal{E}_\mathrm{las}$.
$\hat{\mathcal{E}}_k$ is the value also normalized by $\mathcal{E}_\mathrm{las}$,
similar to $\tilde{\mathcal{E}}^\mathrm{on}_\mathrm{las}$.
$\hat{\mathcal{E}}_k$ is approximately $30\%$ of
$\tilde{\mathcal{E}}^\mathrm{on}_\mathrm{las}$.
}
\label{fig:fig-9}
\end{figure}

Figure \ref{fig:fig-8} showed the amount of laser energy passing over the disk
target, whereas Fig. \ref{fig:fig-9} illustrates the laser energy on the disk
surface.
The blue solid line in Fig. \ref{fig:fig-9} shows the relationship between the
disk diameter and laser energy on a circle of diameter $d$ placed with its plane
facing the laser (i.e., on the initial disk surface),
$\mathcal{E}^\mathrm{on}_\mathrm{las}$, (given by Eq. (\ref{r_ene})), shown as
a ratio to the energy of the initial laser pulse, $\mathcal{E}_\mathrm{las}$,
(i.e., $\tilde{\mathcal{E}}^\mathrm{on}_\mathrm{las}$=$\mathcal{E}^\mathrm{on}_\mathrm{las}/\mathcal{E}_\mathrm{las}$).
This analytical line is also depicted in Fig. \ref{fig:fig-8} (inset),
inverted vertically as
$\tilde{\mathcal{E}}^\mathrm{on}_\mathrm{las}=1-\tilde{\mathcal{E}}^\mathrm{out}_\mathrm{las}$.
$\tilde{\mathcal{E}}^\mathrm{on}_\mathrm{las}$
can also be regarded as being normalized so that its maximum value is $1$.
The red circles show the simulation results of the total kinetic energy,
$\tilde{\mathcal{E}}_k$, of the disk target (i.e., the total kinetic energy of
the electrons and ions) at each disk diameter, and the red line represents
their fitting line.
$\tilde{\mathcal{E}}_k$ and $\tilde{\mathcal{E}}^\mathrm{on}_\mathrm{las}$
are the values normalized to the maximum value of $1$ in each line.
The simulation and analytical curves are in good agreement.
As the laser energy on the disk surface
$\tilde{\mathcal{E}}^\mathrm{on}_\mathrm{las}$ increases, the total kinetic
energy of the entire disk (i.e., of all the ions and electrons),
$\tilde{\mathcal{E}}_k$, also increases,
causing the trends of the two lines to coincide.
Both $\tilde{\mathcal{E}}^\mathrm{on}_\mathrm{las}$ and $\tilde{\mathcal{E}}_k$
increase as the disk diameter increases (implying that it changes from a disk
of small diameter to a foil) until a certain diameter ($d \gtrsim 7\mu$m) is
reached, after which they become constant.
As the disk diameter increases, $\tilde{\mathcal{E}}_k$ increases, but this is
due to an increased number of accelerated ions and electrons rather than the
production of higher-energy ions.
At a diameter of $d = 3\mu$m, where ion energy attains its maximum,
$\tilde{\mathcal{E}}_k$ is only approximately half of its peak value.
This indicates that concentrating the laser energy on a smaller number of ions
and electrons is important to produce high-energy ions.
In comparison with foils, disks selectively impart a greater amount of laser
energy to a smaller number of ions.
To compare the energy values of the two curves in Fig. \ref{fig:fig-9},
$\mathcal{E}_k$ is normalized by the laser pulse energy
$\mathcal{E}_\mathrm{las}$,
as is $\tilde{\mathcal{E}}^\mathrm{on}_\mathrm{las}$, to obtain
$\hat{\mathcal{E}}_k$, and displayed in the inset of Fig. \ref{fig:fig-9}.
The analytical curve (blue line) is the same as in Fig. \ref{fig:fig-9}.
$\hat{\mathcal{E}}_k$ is much smaller than
$\tilde{\mathcal{E}}^\mathrm{on}_\mathrm{las}$, that is, approximately $30\%$ of
$\tilde{\mathcal{E}}^\mathrm{on}_\mathrm{las}$.
This is because, in the simulation, the laser is partly reflected from the disk
surface and partly transmitted through the disk. 

\begin{figure}[tbp]
\includegraphics[clip,width=\hsize,bb=203 53 665 425]{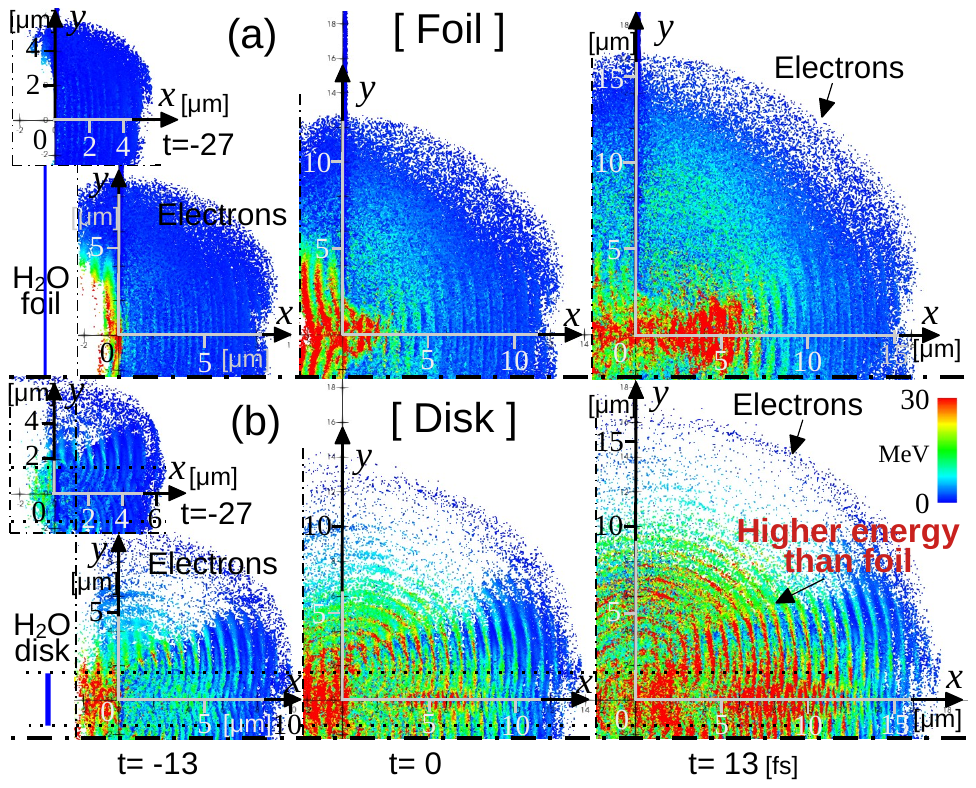}
\caption{
Spatial distribution of the electrons at each time instant in the region near
the target where $x \gtrsim 0$, $y \gtrsim 0$.
2D projection of the section at $z=0$ as viewed along the $z$ axis.
The electrons are colored according to their energy.
The energy of the electrons distributed in the area above ($+y$ side of) the top
of the initial disk target is higher in the disk case than in the foil case.
}
\label{fig:fig-10}
\end{figure}

Figure \ref{fig:fig-10} shows the spatial distribution of the electrons in
regions where $x \gtrsim 0$ and $y \gtrsim 0$ around the initial target
for the (a) foil and (b) disk cases.
The cross section at $z=0$ is shown in 2D.
The $z$ direction is the same as that in Fig. \ref{fig:fig-7}.
The electrons are color coded according to their energy,
with red and blue indicating high and low energies, respectively.
In the case of the disk (b), the top and bottom positions
of the initial disk are denoted by two black dotted lines.
In both cases, the electrons spread farther in the $+x$ and $+y$ directions
(and the $-y$ direction in the $y<0$ region) with time, and the distribution
range of the electrons at each instant is almost the same.
As time passes, the area where the high-energy electrons are distributed
(red area) gradually expands, and this area is wider in the disk case
(b) than in the foil case (a).
Because the laser intensity is the strongest at $y=0$, high-energy electrons
are generated near the $x$ axis in both cases.
At each time, the high-energy electrons are distributed farther
in the $+x$ direction in the disk case than in the foil case.
In the area above the $+y$ side of the top of the initial disk target
($y>1.5\mu$m), i.e., the region above the upper black dotted line in
Fig. \ref{fig:fig-10}(b), the electrons in the disk case have much higher energy
than that in the foil case.
This shows that the electrons pushed out from the disk in the $+y$ direction
are heated by the laser passing through the area beyond the disk.
This is why disks generate high-energy ions more effectively than foils.

\begin{figure}[tbp]
\includegraphics[clip,width=0.9\hsize,bb=33 38 561 403]{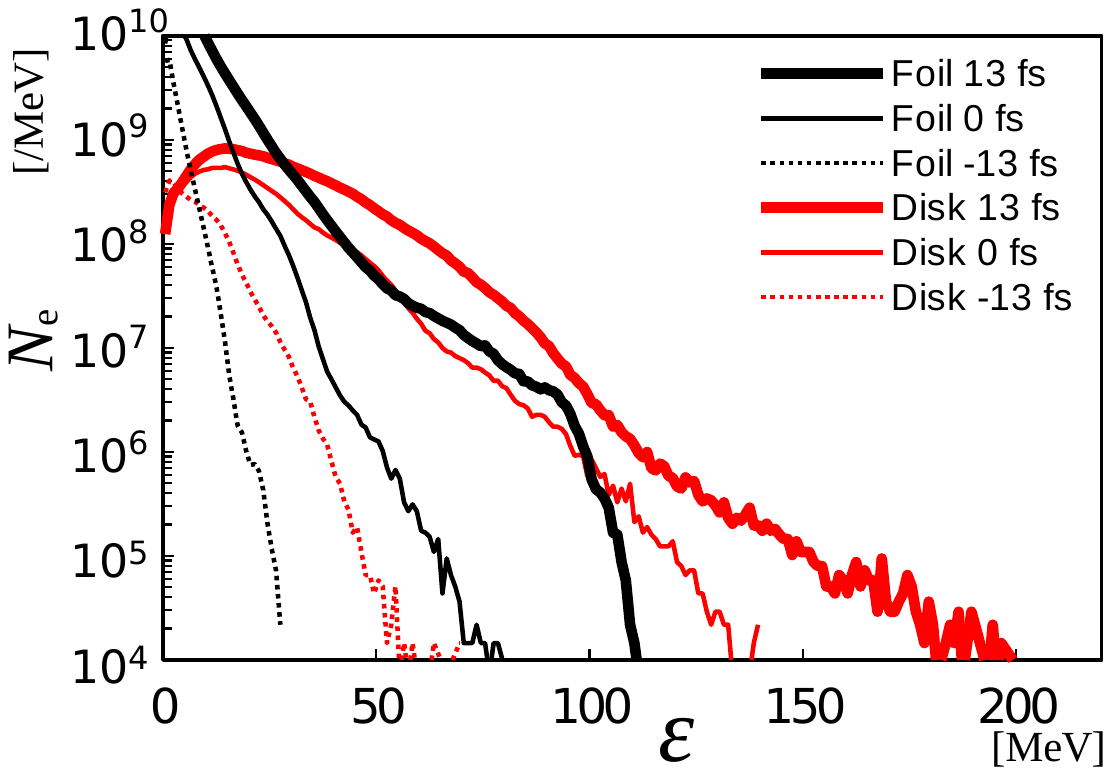}
\caption{
Energy spectra of the electrons in the region $x>0.2\mu$m, $y>1.5\mu$m.
At each instant,
the energy of the electrons is much higher in the disk case than in
the foil case.
Furthermore, the number of electrons in the high-energy region is greater
in the disk case than in the foil case.
}
\label{fig:fig-11}
\end{figure}

As illustrated in Fig. \ref{fig:fig-10}, in the region beyond the circumference
of the disk, the electrons in the disk case have higher energy than that
in the foil case; this trend is also evident in the energy spectrum of the
electrons.
Figure \ref{fig:fig-11} shows the energy spectrum of the electrons within the
region similar to that displayed in Fig. \ref{fig:fig-10}.
It represents the energy spectrum of the electrons present in the region
of $x>0.2\mu$m (the $+x$ side from the surface of the disk target),
$y>1.5\mu$m (the $+y$ side from the top of the disk target), and the entire area
in the $z$ direction.
The results for the foil (red line) and disk (black line) are shown at
$t = -13$ fs (dotted line), $0$ (thin solid line), and $13$ fs
(thick solid line), which are the same time points as depicted in the spatial
distribution in Fig. \ref{fig:fig-10}.
In the low-energy range, the number of electrons in the foil case is much
greater than that in the disk case.
However, in all other energy ranges, the electrons in the disk case are more
numerous than those in the foil case and have much higher energy at all times.
This is because, as shown in Fig. \ref{fig:fig-10}, the electrons that are
pushed out from the disk in the $\pm y$ direction,
which is actually the $+r$ direction in 3D, absorb the energy of the laser
passing outside the disk and are heated.
Moreover, this figure shows that such electrons are numerous.

\begin{figure}[tbp]
\includegraphics[clip,width=0.9\hsize,bb=30 32 562 406]{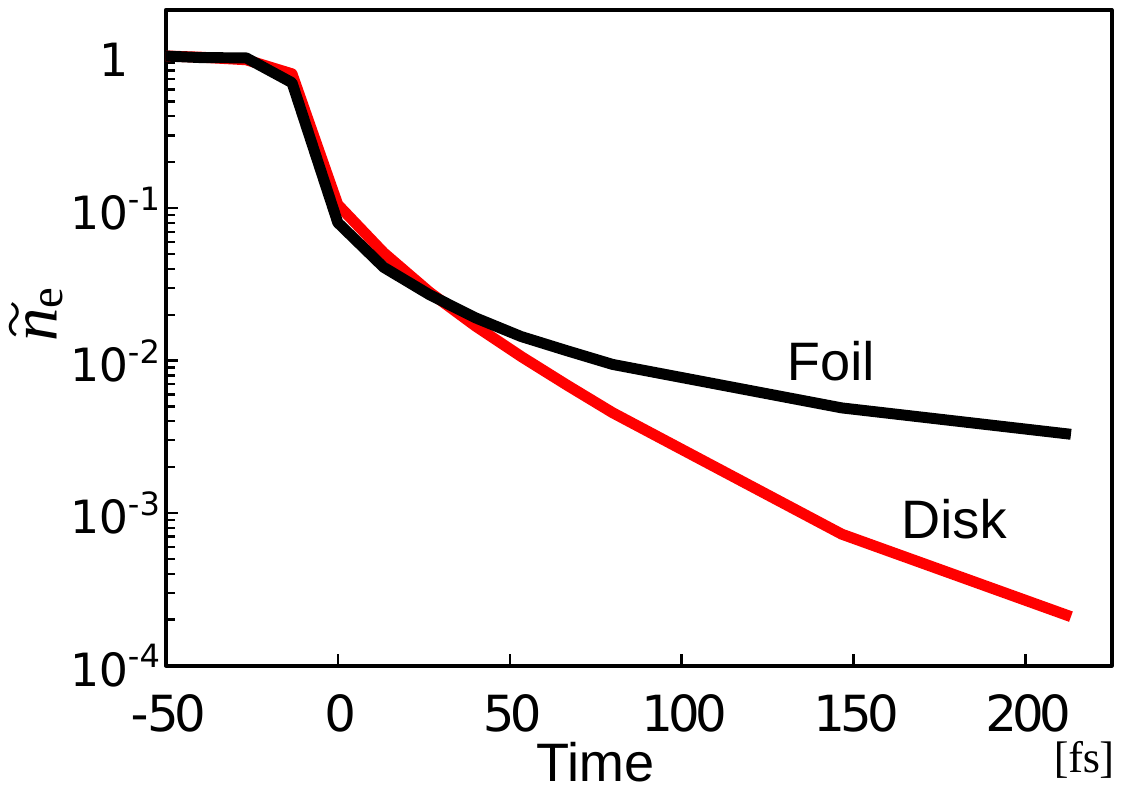}
\caption{
Electron density in the target as a function of time in the foil and disk cases.
In both cases, the number of electrons in this area continues to decrease.
Up to $t \approx 50$ fs, the electron density is almost the same in the foil
and disk cases, but for $t \gtrsim 50$ fs,
it is smaller in the disk case than in the foil case.
This is because the electrons are accelerated more effectively in disk targets
than in foil targets.
}
\label{fig:fig-12}
\end{figure}

Figure \ref{fig:fig-12} shows the change in the number density of electrons
over time in the region within the initial target (the initial disk area,
i.e., the area with diameter = $3\mu$m and thickness = $0.2\mu$m) for both
the foil and disk cases.
The number density, $n_e$, is calculated as $n_e=N_e/V_t$, where $N_e$ is the
number of electrons in this region and $V_t$ is the volume of this region,
and is normalized by the value at the initial time, $n_e(0)$, as
$\tilde{n}_e(t)=n_e(t)/n_e(0)$.
In both cases, the number of electrons in the target decreases rapidly
from around $t=-20$ fs and continues to decrease until the end.
At $t \lesssim 50$ fs, the electron densities in both cases are almost the same,
but at $t \gtrsim 50$ fs, the density in the disk case decreases more than
that in the foil case.
This is because the electrons in the disk case are heated more effectively
than in the foil case, for the same reason as explained above.
However, even in the foil case, the electron density continues to decrease.
Thus, in our simulation, there is no phenomenon wherein the electrons flow into
the laser-irradiated region from the surrounding area,
resulting in an increase in the number of electrons in that region.
At $t=50$ fs, the electron densities in both cases decrease to approximately
$1/100$ of the initial value.
The energy obtained by a proton placed at the center of the surface of a disk
with radius $R$, thickness $\ell$, and charge density $\rho$ is given by
$\mathcal{E}_p=q_e E_0 R$, where $q_e$ is the electron charge,
$E_0=\rho\ell/2\epsilon_0$, and $\epsilon_0$ is the vacuum permittivity.
The difference between the proton energies
$\mathcal{E}_p$ and $\mathcal{E}'_p$ from the disks with
$\tilde{n}_e =0.01$ and $\tilde{n}'_e =0$
(i.e., all the electrons removed), respectively, is
$\mathcal{E}_p/\mathcal{E}'_p = \rho/\rho' = q_e(1-\tilde{n}_e)n_e(0) / q_e n_e(0) = 1-\tilde{n}_e$,
where $\rho$ and $\rho'$ are the charge densities at
$\tilde{n}_e=0.01$ and $\tilde{n}'_e=0$, respectively.
Then, $\mathcal{E}_p/\mathcal{E}'_p$ is 0.99,
which indicates that the values are nearly the same.
Thus, for $\tilde{n}_e<0.01$, the electron density is practically 0 in terms
of the ion acceleration.
Therefore, at $t \gtrsim 50$ fs, the electron density is higher in the foil case
than in the disk case, and at the final instant, $t=213$ fs, the difference is
approximately 10 times greater; however, this difference has a negligible
effect on the ion acceleration.

\begin{figure}[tbp]
\includegraphics[clip,width=\hsize,bb=53 45 514 525]{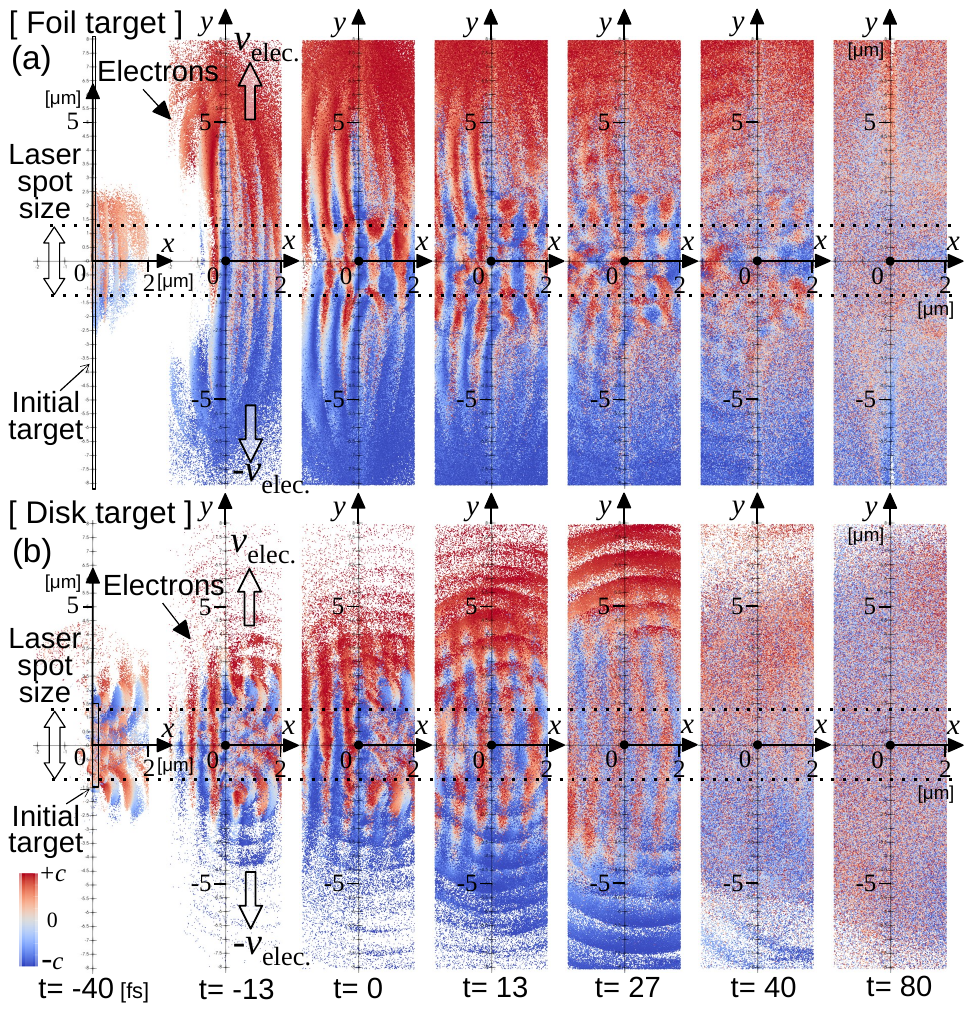}
\caption{
Distribution of electrons near the target.
The electrons are colored according to their velocity in the $y$ direction.
The velocity of the electrons is distributed as red and blue vertical stripes
near the laser-irradiated area, whereas in farther areas in the $y$ direction,
the electrons move outward from the target in the $y$ direction.
}
\label{fig:fig-13}
\end{figure}

Figure \ref{fig:fig-13} shows the distribution of electrons near the target at
each instant, color coded by their $y$ direction velocity.
The cross sections at $z=0$ near the target in the laser-irradiated area
($-8<y<8\mu$m, $-2<x<2\mu$m) for both the foil (a) and disk (b) cases are
presented in 2D.
The $z$ direction is the same as in Fig. \ref{fig:fig-7}.
Red and blue represent electrons moving in the $+y$ (upward in the figure) and
$-y$ (downward in the figure) directions, respectively; the stronger the color,
the higher the speed.
The initial target is shown as a solid black line at $t=-40$ fs, and
the range of the laser spot (FWHM) is indicated by the black dotted line.
For $t \le 27$ fs, the red ($v_y>0$) and blue ($v_y<0$) regions appear as
vertical stripes near the laser-irradiated area (near $y=0$) in both cases.
This pattern arises because the laser electric field in our simulations is
oriented in the $y$ direction, which repeatedly alternates between the upward
and downward directions.
When the laser electric field is oriented in the $-y$ direction (downward),
the electrons are accelerated upward and have a velocity of $v_y>0$ (red); when
it is oriented in the $+y$ direction (upward), they are accelerated downward and
have a velocity of $v_y<0$ (blue).
The red and blue regions appear as vertical stripes, and the distance between
these stripes (from red to red and blue to blue) is approximately the laser
wavelength ($0.8\mu$m).
The distributions of the vertical stripes of the electron velocity are similar
for both the foil and disk cases.
Beyond the laser-irradiated area, owing to the ponderomotive force of the laser
pulse, the electrons move at a higher upward velocity on the $+y$ side and at
a higher downward velocity on the $-y$ side of the laser-irradiated area in
both cases, that is, the electrons move outward from the targets
in the $y$ direction.
As time passes, the separation between the red and blue colors becomes less
distinct, and the color separation is no longer clear at $t=80$ fs.
In addition, in the case of the foil, the electrons initially present in the
region around the laser-irradiated area do not flow into the charged
laser-irradiated area (the area within the laser spot size).

\section{Analytical consideration} \label{theory}

The disk radius that produces the highest-energy protons is investigated
analytically.
Assuming that the laser intensity is given by a Gaussian distribution in
the $y$, $z$, and $t$ directions and the distributions in the $y$ and $z$
directions have the same shape.
Therefore, the intensity is represented by
\begin{equation}
I(y,z,t)=I_0e^{-ay^2}e^{-az^2}e^{-bt^2},
\label{l-shape}
\end{equation}
where $I_0$ is the laser peak intensity (appearing at $y=z=t=0$), and
$a$ and $b$ are parameters that determine the shape of the laser pulse.
At this time, the total laser energy is
\begin{equation}
 \mathcal{E}_\mathrm{las} = \int_{-\infty}^{\infty} \int_{-\infty}^{\infty}
  \int_{-\infty}^{\infty} I(y,z,t)dydzdt = I_0\frac{\pi}{a}\sqrt{\frac{\pi}{b}}.
\end{equation}
The laser energy irradiating onto a disk of radius $R$ with its center at
$y=z=0$ on the $yz-$plane is
\begin{eqnarray}
 \mathcal{E}_\mathrm{R}
   &=& \int_{-\infty}^{\infty}\int_{A}I(y,z,t)dAdt \nonumber \\
   &=& I_0\int_{0}^R\int_{0}^{2\pi}e^{-ar^2}rd\theta dr
       \int_{-\infty}^{\infty}e^{-bt^2}dt \nonumber \\
   &=& I_0\frac{\pi}{a}(1-e^{-aR^2})\sqrt{\frac{\pi}{b}},
\label{dsk_e}
\end{eqnarray}
where $A$ is the region of the disk surface and $r^2=y^2+z^2$.
The ratio of the laser energy on a disk target of radius $R$ to the total laser
energy is
\begin{equation}
 \tilde{\mathcal{E}}_\mathrm{R} =
   \frac{\mathcal{E}_\mathrm{R}}{\mathcal{E}_\mathrm{las}}
   =1-e^{-aR^2}.
\label{r_ene}
\end{equation}
The greater the value of $R$, the greater the laser energy received
by the entire disk target.
The average energy per unit volume of the disk target is
\begin{equation}
 \tilde{u} = \frac{\tilde{\mathcal{E}}_\mathrm{R}}{V_d}
  = \frac{1-e^{-aR^2}}{\pi R^2 \ell},
\label{u_ene}
\end{equation}
where $V_d=\pi R^2 \ell$ is the disk volume and $\ell$ is the disk thickness.
Because the charge density $\rho$ of the charged disk generated by
laser irradiation can be considered to increase with $\tilde{u}$,
\begin{equation}
 \rho = k \tilde{u} = \frac{k(1-e^{-aR^2})}{\pi R^2 \ell},
\label{rho}
\end{equation}
where $k$ is a proportionality constant.

The $x$ component of the electric field of a positively charged thin disk is
\begin{equation}
E_x(x)=\frac{\rho \ell}{2\epsilon_0} \left(1-\frac{x}{\sqrt{x^{2}+R^{2}}}
       \right),
\label{exx}
\end{equation}
where $\epsilon_0$ is the vacuum permittivity.
We assume that the $x$ axis is perpendicular to the disk surface placed at the
disk center.
The energy gain of protons in this static electric field is
$\mathcal{E}_p=\int_0^\infty q_eE_x(x)dx = q_e R \rho \ell/2\epsilon_0$,
where $q_e$ is the electron charge.
Substituting the $\rho$ obtained above into this equation gives
\begin{equation}
\mathcal{E}_p = \frac{q_e k}{2\pi\epsilon_0} \cdot \frac{1-e^{-aR^2}}{R}.
\label{p_ene}
\end{equation}
The obtained proton energy $\mathcal{E}_p$ is expressed as a function of
the disk radius $R$.

Let us find the value of $R$ that maximizes $\mathcal{E}_p$, i.e., find $R$ that
satisfies $\partial \mathcal{E}_p / \partial R = 0$.
\begin{eqnarray}
\frac{\partial\mathcal{E}_p}{\partial R} 
 &=& \frac{q_e k}{2\pi\epsilon_0}
 \cdot \frac{\partial}{\partial R} \left(\frac{1-e^{-aR^2}}{R} \right)
 	\nonumber \\
 &=& \frac{q_e k}{2\pi\epsilon_0}
 \left[ (2a+\frac{1}{R^2}) e^{-aR^2} - \frac{1}{R^2} \right].
\label{de_dr}
\end{eqnarray}
From $\partial \mathcal{E}_p / \partial R = 0$, we obtain
\begin{equation}
 e^{aR^2} - 2aR^2 -1 = 0.
\label{dedr0}
\end{equation}

Set $aR^2=\xi$, and solve $e^{\xi}-2\xi-1=0$ for $\xi$.
The derivative of this function is 0 at $\xi=\log2$, negative for $\xi<\log2$,
and positive for $\xi>\log2$; the function value at $\xi=\log2$ is $<0$.
Therefore, there are two solutions.
One solution is $\xi=0$, which can be easily deduced.
In this solution, we obtain $R=0$, but this does not meet our requirement.
Because this equation is a transcendental equation, it is solved numerically
using Newton's method to give $\xi=1.256$.
Considering the relationship $a=\log2/r_\mathrm{las}^2$,
\begin{equation}
R = \sqrt{ \frac{\xi}{\log2} } \cdot r_\mathrm{las} \approx 1.3 ~r_\mathrm{las},
\label{rrr}
\end{equation}
where $r_\mathrm{las}$ is the laser spot radius (HWHM).
Thus, the radius $R$ of the disk target that produces the maximum
proton energy is $1.3$ times the laser spot radius (HWHM),
$r_\mathrm{las}$.
This theoretical value is in good agreement with the simulation
value $d/D_L = 1.2$, shown above. 

\section{Conclusions} \label{con}

Laser ion acceleration using H$_2$O foil and disk targets was investigated
using 3D PIC simulations.
Using H$_2$O, a hydrogen-rich material, enables the generation of 200 MeV-class
protons at
$\mathcal{E}_\mathrm{las}=25$ J, $P=0.8$ PW, and $I=1 \times 10^{22}$ W/cm$^2$.
To generate high-energy ions, an optimum thickness suitable for the laser
conditions exists for the foils and an optimum thickness and diameter for the
disks.
The generated ion energy diminishes as it strays from the optimal value.
Mass-limited targets, such as disks, produce higher-energy ions compared to
foils.
This is because, in the case of a disk, the electrons expelled from the disk
are effectively heated by the laser passing through the area beyond the disk.
This means that a part of the laser pulse incident on the area beyond the disk
is absorbed into the electrons of the initial disk, which has a narrower area,
i.e., it is the same as focusing the laser.
That is, mass limited targets, such as disks, have the effect equivalent to
laser focusing occurring during the acceleration process.
That means, the laser "intensity" increases.
Therefore, the optimal disk thickness for generating high-energy ions exceeds
that of foils, being twice as thick in our simulations.

The optimal disk diameter, which makes this laser-focusing effect noticeable,
is $\sim$1.2 times the laser spot diameter.
Under the conditions of our study, the target thickness that yields maximum
energy ranges from 0.1 to 0.2 $\mu$m.
Soap films, primarily composed of water, have thicknesses ranging from submicron
to a few microns; thus, it is possible to prepare a 0.1--0.2 $\mu$m-thick
H$_2$O foil target using soap films.
In addition, in our simulations, disks float in space with one surface facing
the direction of the laser.
It may be difficult to create such a situation, but it is possible to
approximate this state by placing a small dot of water
(or removing excess water around a small circular area) on an ultrathin film
that is only a few nanometers thick (e.g., graphene or diamond-like carbon).
This is because, in this study, the disk thickness ($\sim 200$ nm) is more than
100 times greater than the thin film to which the disk is attached;
therefore the effect of the thin film is very small. 

One of the characteristics of water is its transparency.
A significant issue with thin targets in laser ion acceleration is that
the pre-pulse present before the main laser pulse destroys the target before
the arrival of the main pulse.
Owing to the transparency of water, the pre-pulse, which has a lower intensity
than the main pulse, can pass through the water target without destroying it,
and the interaction between the target and the main pulse can be achieved.

\section*{Acknowledgments}
I thank S. V. Bulanov, T. Zh. Esirkepov, R. Hajima, M. Kando, J. Koga
and H. Kotaki for their valuable discussions.
This research was conducted with the supercomputer HPE SGI8600 in
the National Institutes for Quantum Science and Technology.



\end{document}